\documentclass[preprint,12pt]{elsarticle}

\usepackage{amssymb}

\usepackage{threeparttable} 

\journal{Physica A}

\begin{document}

\begin{frontmatter}



\title{Testability of evolutionary game dynamics
based on experimental economics data}


\author{}

\address{}
\author[label1]{Yijia Wang}
\author[label1]{Xiaojie Chen}
\author[label2]{Zhijian Wang}
\address[label1]{School of Mathematical Sciences, University of Electronic
Science and Technology of China, Chengdu 611731, China}
\address[label2]{Experimental Social Science Laboratory, Zhejiang University,
Hangzhou 310058, China}

\begin{abstract}
Understanding the dynamic processes of a real game system requires an appropriate dynamics model, and rigorously testing a dynamics model is nontrivial.
In our methodological research, we develop an approach to testing the validity of game dynamics models that considers the dynamic patterns of angular momentum and speed as measurement variables. Using Rock-Paper-Scissors (RPS) games as an example, we illustrate the geometric patterns in the experiment data. We then derive the related theoretical patterns from a series of typical dynamics models. By testing the goodness-of-fit between the experimental and theoretical patterns, we show that the validity of these models can be evaluated quantitatively.
Our approach establishes a link between dynamics models and experimental systems, which is, to the best of our knowledge, the most effective and rigorous strategy for ascertaining the testability of evolutionary game dynamics models.

\end{abstract}


\end{frontmatter}







 \newpage
\tableofcontents
 \newpage
\section{Introduction}
\subsection{Research question}
\begin{figure}[ht]
  \begin{center}
    \begin{flushleft}
    ~~~~~~~~~~~~~~~~~~~~ Payoff Matrix   ~~~~ Replicator  ~~~~~~ Projection  ~~~~~\\
    ~~~~~~~~~~~~~~~~~~~~ ~~~~~~~~~~~~ ~~~~~~~~~~  Dynamics ~~~~~~~  Dynamics ~~~~~
\\\end{flushleft}
~\\
    \includegraphics[angle=0,width=0.7\textwidth]{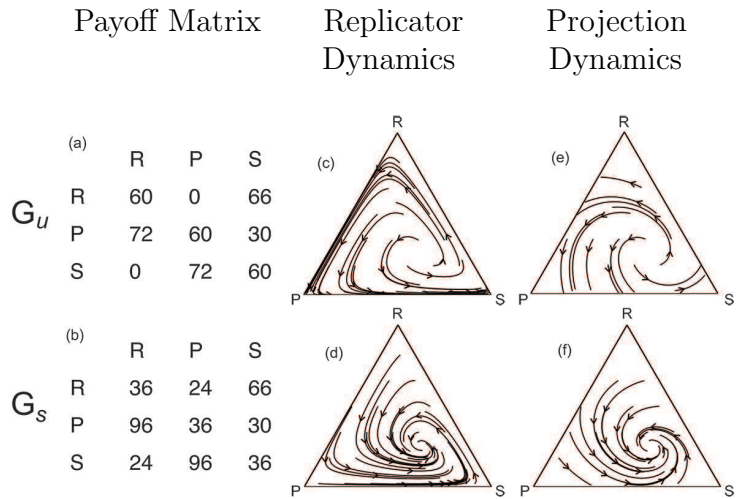}
  \end{center}
  \caption{
    \textbf{Evolutionary trajectories of the two RPS games derived from two models.}
    (a) and (b) depict the payoff matrices for the RPS games used as examples in Ref. \cite{Friedman2014} and this study, and are unstable (denoted by $G_u$) and stable (denoted by $G_s$) RPS games, respectively. Theoretically, the rest points of $G_u$ and $G_s$ are identical in (1/4, 1/4, 1/2) in the state space (also called the simplex).
    In (c-f), the black lines and arrows indicate evolutionary trajectories. In (c) and (d), the evolutionary trajectories are derived from the Replicator dynamics model using $G_u$ and $G_s$, respectively. Similarly, in (e) and (f), the results come from the Projection dynamics model.
  \label{fig:fig1}
  }
\end{figure}

Evolutionary game theory, rooted in classical game theory \cite{VonNeumann1944,myerson1997game} and evolutionary theory \cite{darwin1859origin}, has been widely used to study the dynamical behaviors of game systems
\cite{Smith1982,Levin2009Games,Weibull1997,HofbauerSigmund1998,Skyrms2014Social,
Samuelson2002,Skyrms2014In,Friedman1998Rev,nowak2006evolutionary,Sandholm2011,frey2010evolutionary}.
Since the Replicator dynamics model was first proposed \cite{Taylor1978},
a substantial number of evolutionary game dynamics models have been developed (e.g., \cite{Sandholm2011}). These dynamics models can be classified by their update protocols \cite{Sandholm2011} or geometric properties \cite{sandholm201603}, and can thus produce rich theoretical evolutionary dynamics.
Each model has its own quantitative predictions.
For example, as shown in Fig. \ref{fig:fig1}, two Rock-Paper-Scissors (RPS) games with identical rest points (or Nash equilibria, in classical game theory) have different trajectories induced by the same model. However, for the same RPS game, the trajectories induced by different models are obviously different. These findings have now become standard textbook content on evolutionary dynamics
\cite{Smith1982,Weibull1997,HofbauerSigmund1998,nowak2006evolutionary,Sandholm2011}.

Naturally, the reality of evolutionary dynamics models have been widely considered \cite{Sin96,pryke2006red,Ker02,Levy2015Quantitative}. In recent decades, social scientists have continued to investigate and improve upon the commonalities between models and experiments \cite{Crawford1991,Miller1991Can,Friedman1998Rev,Plott2001Equilibrium,Plott2008,Friedman1996,Friedman2014,Wang2014Social,Nowak2015,Friedman2016,Wangxu2014}. However, there is still a significant gap between evolutionary outcomes from models and empirical results.
Without loss of generality, the gap can be seen in representative RPS game experiments \cite{Wang2014Social,Wangxu2014,Friedman2014,Nowak2015,friedman2010tasp,Zhou2016}. Scientists have clearly illustrated how to distinguish various games with models in experiments \cite{friedman2010tasp,Wang2014Social,Wangxu2014,Friedman2014,Nowak2015};
however, distinguishing various models associated with games in experiments has only rarely been achieved. For a given game experiment, it seems that the dynamics model could be arbitrarily chosen because various models have similar expectations
regarding existing observations, which is unsatisfying. Hence, rigorous testing of game dynamics models has been an open question.

\subsection{Logic of our approach}
In this methodological study, we aim to answer this question through a novel approach inspired by two recent advances in game experiments. The first concerns measurements. In game experiments, by considering the time reversal symmetry in high stochastic trajectories of social state motions, the observations of deterministic motions (e.g., cycle frequency \cite{XZW2013,Xu2014,Zhou2016}, cyclic motion vector field \cite{Wang201009,XuWang2011ICCS,xu2012test}, and cycle counting index \cite{Friedman2014}) have been quantified, which led us to explore new measurements of dynamic patterns.
The second development is in the domain of experimental technology, which has enabled the realization of continuous time experiments from which sufficiently long trajectories can be harvested \cite{Friedman2011Separating,Friedman2014}. The continuous-time, continuous-strategy, and instantaneous treatment of the two RPS games shown in Fig. \ref{fig:fig1} \cite{Friedman2014} is an exemplar of such experiments. Without loss of generality, the data from this experiment can be employed to demonstrate our approach.

This paper is organized as follows. Section 2 describes the two measurements for
dynamical observations: angular momentum $L$ and speed $S$. In Section 3, using RPS games experiments data\cite{Friedman2014}, we demonstrate experimental dynamical patterns.
In Section 4, we derive the theoretical dynamical patterns from a series of typical dynamics models specified by the RPS game payoff matrix. We then, in Section 5, test the goodness-of-fit of the theoretical and experimental patterns.
From this procedure, our proposed approach can distinguish which dynamics models should be considered as candidates for describing the dynamical behaviors of a given experimental system. In Section 7, the advantages of this approach, as well as related literature and further research questions, are discussed.

\section{Measurements for dynamical patterns}
\subsection{Time series, evolutionary trajectory, and velocity}
\begin{figure}[ht]
  \begin{center}
  (a)\\
    \includegraphics[angle=0,width=0.9\textwidth]{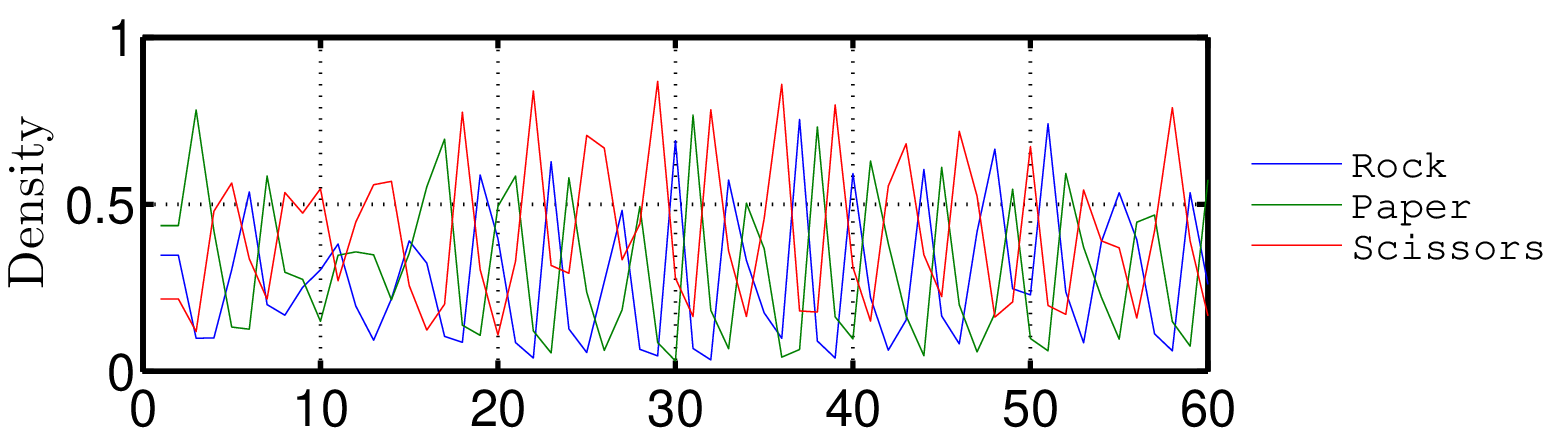} \\
 ~\\
 (b)~~~~\\
    \includegraphics[angle=0,width=0.3\textwidth]{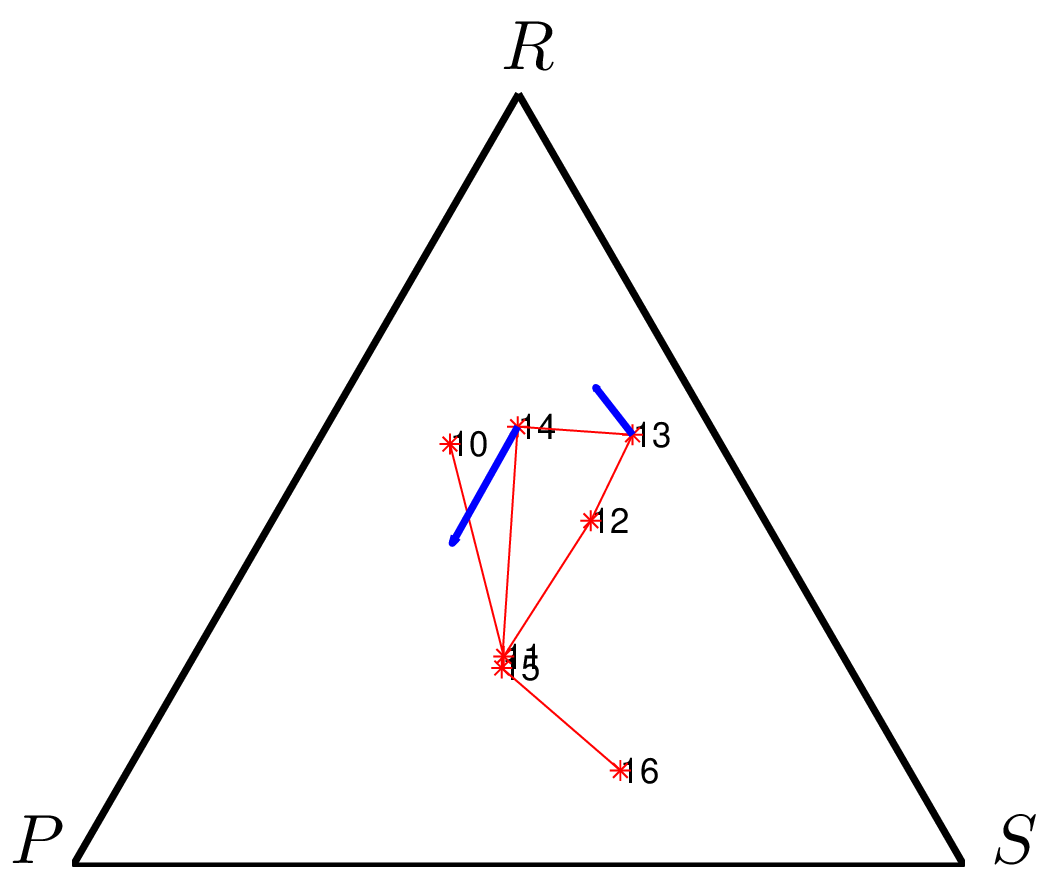} ~~~~
  \end{center}
  \caption{Time series, evolutionary trajectory, and dynamical observation. (a) is a sample time series of 60-second records from the RPS game experiments. (b) depicts the time series (from the seconds 10--16 in (a)) in the phase space of the RPS game. The red segments indicate the evolutionary trajectory and the blue arrow indicates instantaneous velocity.  \label{fig:SI_TimeSeries}}
\end{figure}

Without loss of generality, in a real-time RPS game experiment system, we can obtain a time series, as shown in Fig.\ref{fig:SI_TimeSeries} (a), indicating strategy fractions as a function of time. We can use a strategy vector given by $[P_R, P_P, P_S]$ to represent the fractions of Rock, Paper, and Scissors, respectively, at each time step. Correspondingly, the strategy vector at each time step can be plotted as a state point in the simplex (state space), and the state points for all strategy vector values form the evolutionary trajectory in experiments, as shown in Fig. 2(b).
Based on the evolutionary trajectory in the phase space, we can introduce two measurements at state \textbf{x}, speed $S_\textbf{x}$ and angular momentum $\textbf{L}_{\textbf{x}}$, as follows. \\
\subsection{Speed $S$ at state \textbf{x}}
We assume that the state of strategy vector \textbf{x} at time $t$ is \textbf{x}$(t)$, which is written as [$P_{R},P_{P},P_{S}$]. Accordingly, the states at $t$-1 and $t$+1 are \textbf{x}$(t-1)$ and \textbf{x}$(t+1)$, respectively. We then define a jump-out transition for state \textbf{x}$(t)$ as $\textbf{x}^{+}=\textbf{x}(t+1)-\textbf{x}(t)$. Similarly, a jump-in  transition for state \textbf{x}$(t)$ is
$\textbf{x}^{-}=\textbf{x}(t)-\textbf{x}(t-1)$. Thus, an observation of instantaneous velocity $\textbf{v}_{\textbf{x}}$ at state \textbf{x}$(t)$ can be defined as \cite{XuWang2011ICCS}
 \begin{equation}\label{eq:phiV0805}
    \textbf{v}_{\textbf{x}}=\frac{\textbf{x}^{+}+\textbf{x}^{-}}{2\Delta t}=\frac{\textbf{x}(t+1)-\textbf{x}(t-1)}{2\Delta t},
 \end{equation}
where $\Delta t$ is the time interval, which is set to one in the two RPS experiments. In Figure \ref{fig:SI_TimeSeries} (b), two examples of instantaneous velocity are illustrated. The corresponding average instantaneous velocity value observed at \textbf{x} is \cite{XuWang2011ICCS}, 
$\bar{\textbf{v}}_{\textbf{x}}=[\bar{v}_{R},\bar{v}_{P}, \bar{v}_{S}]$,
where $\bar{v}_{R}$, $\bar{v}_{P}$, and $\bar{v}_{S}$ are the average velocity components at state \textbf{x}. This measurement of average velocity is of time-reversal asymmetry and describes the deterministic motion observed \cite{XZW2013,XuWang2011ICCS}.

To clearly compare the speed values in various models and in experiments, we further define the magnitude of average velocity $\bar{\textbf{v}}_{\textbf{x}}$ at state \textbf{x} as
\begin{equation}\label{eq:S0804}    S_{\textbf{x}}=||\bar{\textbf{v}}_{\textbf{x}}||=\sqrt{\bar{v}_{R}^{2}+\bar{v}_{P}^{2}+\bar{v}_{S}^{2}}.
\end{equation}
\subsection{Angular momentum $\textbf{L}$ at state \textbf{x}}
Based on the definition of velocity, we further define angular momentum $\textbf{L}_{\textbf{x}}$ at state \textbf{x}
as
  \begin{equation}\label{eq:L0804}
     \textbf{L}_{\textbf{x}}=(\textbf{x}-\textbf{O}) \times \bar{\textbf{v}}_{\textbf{x}},
 \end{equation}
where $\times$ indicates the cross-product of the two vectors and ${\textbf{O}}$ is the state vector for the Nash equilibrium strategy in the state space. Correspondingly, $\textbf{L}_{\textbf{x}}=[L_{R}, L_{P}, L_{S}]$, where $L_{R}$, $L_{P}$, and $L_{S}$ are the angular momentum components at state \textbf{x}$(t)$ in the state space (see Figure S2 in
Supplementary Information).
We point out that per the definition of angular momentum, all the angular momentum vectors in the state space should be parallel with the vector $[1, 1, 1]$, which means that the values of the three components are identical. Thus, we can directly choose one component of the vector for comparing angular momentums. For simplicity, we choose $L_R$, define $L=L_R$, and then compare the $L$ values of angular momentums in various models and experiments. Note that the $L$ value for one angular momentum could be negative or positive because the direction of vector $\textbf{L}_{\textbf{x}}$ could be the opposite or the same as that of vector $[1, 1, 1]$. Further, this measurement is also of time reversal asymmetry and describes the deterministic motion observed \cite{XZW2013,XuWang2011ICCS}.


\section{Experimental dynamic patterns}

\subsection{Data}
To illustrate the dynamical patterns in real systems, we use data from RPS game experiments as an example. Eight human subjects participated in the experiments, playing RPS against the other 7 subjects. The experiments are of continuous-time and continuous-strategy instantaneous treatments \cite{Friedman2014}. In the experiments, the strategies used were simply recorded each second in real time; the strategies used by each subject were instantaneously known to all 8 subjects. The payoff matrices are the game parameters controlled by the experimenters, which are shown in Fig. 1 (a) and (b), and respectively represent the unstable and stable RPS games. There are 6300- (5400-)second records from the unstable (stable) RPS game experiments used in our study (For more details, see Section S2.1 and Figure S1 in Supplementary Information). We used these data to illustrate the experimental dynamical patterns employed to test the dynamics models.

\subsection{Method}
By employing the measurement of $\textbf{L}_{\textbf{x}}$ and $S_{\textbf{x}}$ shown in Equation \ref{eq:S0804} and Equation \ref{eq:L0804}, respectively, and using the time series from the two RPS game experiments, we can obtain the experimental dynamical patterns.

\subsection{Results}
Figures \ref{fig:fig2} (i--l) illustrate the experimental dynamic patterns of the angular momentum $L$ and speed $S$ of the two RPS games. In these four figures, the blue (red) color corresponds to the \emph{relative} low (high) values of $L$ or $S$. To present these four figures, we have separated the state space into discrete counterparts, herein called patches or cells, with resolution of 0.050 (denoted by $d$, see Appendix A for its definition. The result of setting the resolution to 0.025 and 0.100 are similar and are shown in Section S3 in Supplementary Information). To our knowledge, in existing literature relating to real systems in evolutionary game theory, such high-precision empirical dynamic patterns have not yet been seen.

Corresponding to the four theoretical results mentioned above, we list four experimental results in Fig. \ref{fig:fig2} and compare them with existing literature \cite{Friedman2014}.
\begin{itemize}
\item First, in each simplex, the observed $L$ and $S$ at state points around the rest point $(\frac{1}{4},\frac{1}{4},\frac{1}{2})$ are minimal and are close to zero. These empirical observations meet the predictions of the nonparametric models well; none of the nonparametric models can be rejected (or excluded) by this empirical result.
\item Second, in all patches in the simplex, the observed $L$ values are not negative in the simplex globally. (In the high-resolution case, we have observed negative values in some patches, though the number of patches is very small and the negative values are very close 0, which can be regarded as noise and ignored in this case study). Thus, with respect to the rest point, the direction of the average motion of social state rotation is counter-clockwise. These empirical observations also meet the expectations of the nonparametric models.
%
\item Third, in the experimental patterns, the numbers of white (blank) patches in the unstable RPS game (Fig. \ref{fig:fig2} (i,j)) are significantly less than those in the stable RPS game (Fig. 2 (k,l)). This means that the stationary distribution (time average of the trajectory distribution) of the unstable RPS game is closer to the simplex edge. This result agrees with the theoretical expectation mentioned above, and with previous results \cite{Nowak2015,Friedman2014}.
%
\item Fourth, for both RPS games, the values of the observed $L$ and $S$ are larger in the patches closer to the edge than those closer to the rest point. That is, the deterministic motions are faster in the social state close to the simplex edge than those close to the rest point. This result was previously unknown; moreover, such clear and high-precision empirical patterns have not been seen \cite{Wang2014Social,Wangxu2014,Nowak2015,Friedman2014}. We use the observed values in each patch to evaluate the theoretical models.
\end{itemize}

The first three experimental results above correspond well to model predictions, but these results provide little help in distinguishing models. To distinguish the models, our approach mainly depends on the fourth result.


\section{Theoretical dynamic patterns}
\subsection{Theoretical Models}
To test whether a dynamics model meets the experimental dynamical patterns, we must deduce the related dynamical observations in the model.
In this study, we evaluate 15 typical evolutionary game dynamics models, which have been clearly summarized and extensively explained in \cite{Sandholm2011,sandholm2007} and its software suite. The models, listed in Table 1, can be classified into two classes: nonparametric and parametric (for which parameters are shown in parentheses following the model name).
Each of the 15 dynamics models has its own mechanism (update rule), and can be explicitly presented as a set of differential equations. For readers who are not familiar with evolutionary game dynamics models, we use two of the 15 models as examples. One example of a dynamics model is the Replicator dynamics model, which can be presented as
\begin{equation}\label{eq:repli}
  \dot{x_i} = x_i (E_{x_i} - \bar{E_x}),
\end{equation}
in which $x_i$ is the density of $i$-strategy, $E_{x_i}$ is the payoff of the $i$-strategists at state $x$, and $\bar{E}$ is the weighted average of the payoff of the population. That is, the density growth rate (update rule) is based on the payoff difference. Another model is the Projection dynamics model (see p. 199 in \cite{Sandholm2011}), in which the growth rate (velocity) of $i$-strategy can be presented as
\begin{equation}\label{eq:projc}
  \dot{x_i} =  E_{x_i} - \tilde{E_x},
\end{equation}
in which $E_{x_i}$ is the payoff of the $i$-strategy population at state $x$ and $\tilde{E_x}$ the unweighted average of the payoff of the population. Importantly, at state \textbf{x} in the state space, the theoretical velocity value is depicted by the differential equations.

\subsection{Method}
With the theoretical velocity at state ${\textbf{x}}$, referring to Equation (\ref{eq:S0804}), the theoretical pattern of speed $S_{\textbf{x}}$ can be obtained; at the same time, by referring to Equation (\ref{eq:L0804}), the theoretical pattern of angular momentum $\textbf{L}_{\textbf{x}}$ can be obtained. Thus, for each of the 15 models, we can obtain the theoretical patterns of $S_{\textbf{x}}$ and $\textbf{L}_{\textbf{x}}$.

\subsection{Results}
Figures 2 (a--h) illustrate the theoretical dynamic patterns of the angular
momentum $L$ and speed $S$ for the stable and unstable RPS games, as derived from the Replicator and Projection dynamics models.
For more theoretical patterns for the models, see Section S4 in Supplementary Information. Among the patterns of the 15 models, four items characterizing the theoretical results, which correspond to the experimental patterns, are listed as follows:
\begin{itemize}
\item First, the angular momentum $L$ and speed $S$ are minimal and close to zero for the state surrounding the rest point $(\frac{1}{4},\frac{1}{4},\frac{1}{2})$. This result is widespread among the nonparametric models, which is meaningful because if this result significantly deviates from the experimental result, all nonparametric models must be rejected and cannot be valid for the given experiments.

\item Second, the expected values of angular momentum $L$ are not negative in the full simplex. Therefore, there should exist counter-clockwise cycles with respect to the rest point in the simplex.

\item Third, in all the dynamics models tested in this study, for the unstable RPS game, the evolutionary trajectories have outward spirals and close in on the simplex edge. In contrast, for the stable RPS game, the evolutionary trajectories have inward spirals and are closing in on the rest point.

\item Fourth, more importantly, the differences between the dynamic patterns derived from different dynamics models are obvious. These differences imply that not every model can be valid for a given experimental system, because the experimental result is unique; only a model in which the pattern matches the experimental pattern well can be valid for the given experiment.
\end{itemize}

We can see that, referring to the first three items, the models usually have the same results. Hence, we cannot use the first three items results to distinguish models.

\section{Model Evaluation}
\subsection{Method}
Quantitatively, we use two coefficients (Pearson correlation coefficient $\rho$ and coefficient of determination $R^2$) to evaluate the validity of a model by comparing the experimental and theoretical dynamical patterns. For a given observation, ideally, $\rho$ and $R^2$ should be 1 (the maximum). The larger the value of $\rho$, the more appropriate the model. Alternatively, if $\rho$ and $R^2$ are close to 0 or are negative, the model is not appropriate for the experimental system. (For details, see \ref{Ap:rhoR2}.)

To illustrate the logic of our approach to evaluating models, we first provide a simple example. Figure ~3 demonstrates how to obtain coefficients $\rho$ and $R^2$ and how to compare the performance of two models. 
Figure~\ref{fig:fig3} (c) illustrates the $\rho$ and $R^2$ of observations $S$ and $L$, which are obtained from Fig.~3 (a) and (b) for the Replicator and Projection dynamics models, respectively.
As illustrated in Fig.~\ref{fig:fig3} (c), with larger $\rho$ and $R^2$ values, the Projection dynamics model is more valid and performs better than the Replicator dynamics model.
%
%
%
\subsection{Results}
For each of the 15 models, the goodness-of-fit ($\rho$ and $R^2$) of the model and experiments are summarized in Table~\ref{tab:Table1}. Here, we show that the validity of a model can be quantified.

With Table~\ref{tab:Table1}, model comparisons can be realized. Table \ref{tab:Table2} reports the statistical results of pairwise comparison of the $15$ models.
(For details of the statistical methods used in these comparisons, see \ref{Ap:rhoR2}.) In each cell in Table \ref{tab:Table2}, the statistic index $+1$ ($-1$) indicates that the model in that row performs significantly better (worse) than that in the column ($p < 0.05$, see Section S6.3.2 in Supplementary Information for $p$-values), while the statistic index $0$ indicates similar performance ($p>0.05$).
Then, for each model, we can obtain its score by adding the statistic indices as shown in the last column in Table \ref{tab:Table2}.
\subsubsection{Explanation of Results}
Table \ref{tab:Table2} can be explained with examples. An example result is that, among the 6 nonparametric models, the Replicator, BR, and MSReplicator dynamics models performed the worst, indicating that these models are not fit for interpreting the experimental system, whereas the Projection dynamics model performs the best. Another example result is that, among the Logit dynamics models tested in this study, the model with parameter 10 performs the best. These examples indicate that the validity of dynamics models can be evaluated and compared quantitatively.

\subsubsection{Robustness of Results}
The results shown in Table \ref{tab:Table2} are robust to various statistical methods (e.g., Student's t-test) and changing resolutions or cut-off counts (see Section S6 in Supplementary Information for more details).
%
The main results in Table \ref{tab:Table2} remain unchanged if we choose only one of the two RPS game experiments, instead of using both.
This means that (1) the validity of a model is independent of the game details. Such a result is common in the interplay between experiments and models in classical game theory \cite{selten2008,RothErev2007}.
(2) Our approach seems efficient in evaluating the validity of various models with one game experiment.

\begin{figure}
\centering
\includegraphics[width=0.95\textwidth]{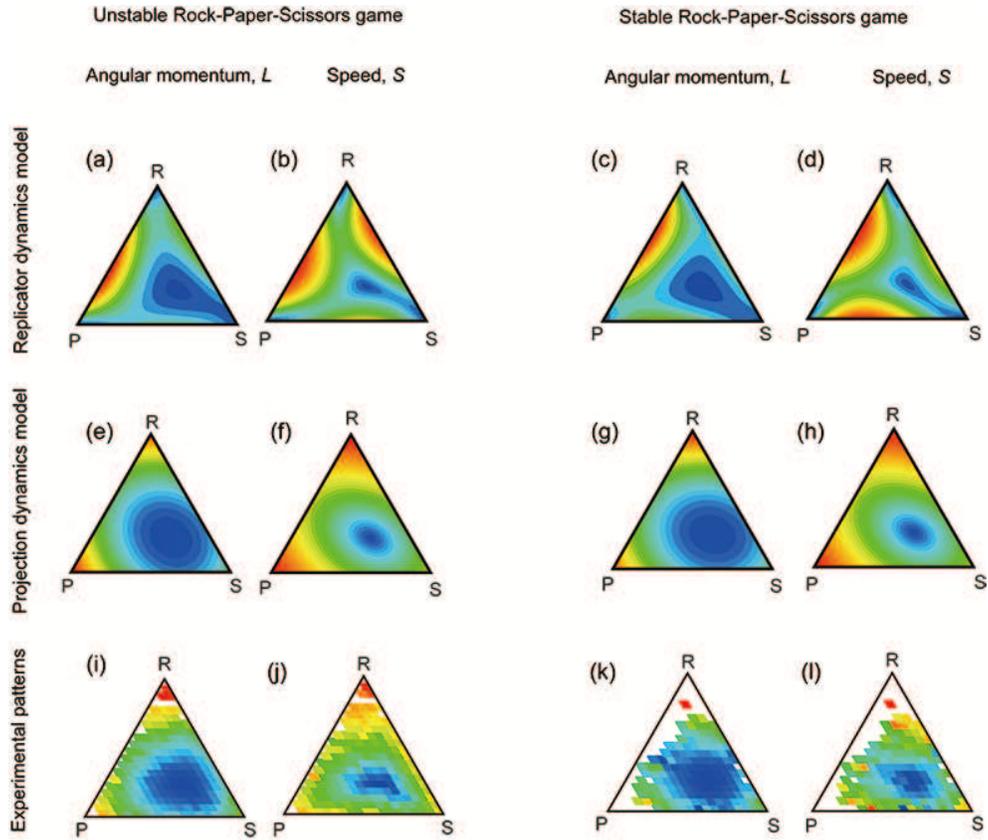}
\caption{{\bf Experimental and theoretical patterns for $L$ and $S$ of the two RPS games.} ($a$--$d$) and ($e$--$h$) depict the theoretical patterns derived from the Replicator and Projection dynamics models, respectively. ($i$--$l$) depict the experimental patterns, which are obtained for the resolution $d=0.050$.
Dark blue corresponds to the lowest value (0, in this case study). For each subfigure, the brightest red indicates the relative highest value of that subfigure. The white patches indicate that no observation was obtained in those patches in the experiments because the evolutionary trajectories could have never visited these patches.
}\label{fig:fig2}
\end{figure}

\begin{figure}
\centering
\includegraphics[width=1\textwidth]{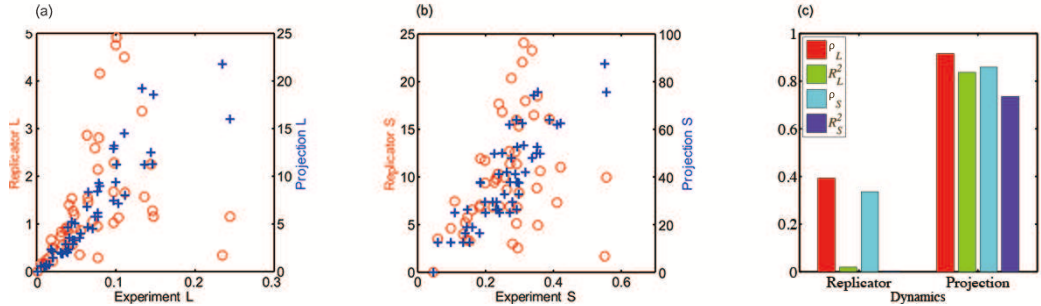}
\caption{{\bf An example comparison of two models.} The theoretical and experimental observations of the same patches in the simplex of the unstable RPS game (resolution $d=0.100$). (a) shows the results of $L$, in which the red circles (blue crosses) depict the experimental versus the theoretical values obtained from the Replicator (Projection) dynamics model. (b) shows the results of $S$. (c) compares the $\rho$ and $R^2$ values of the scatting.
} \label{fig:fig3}
\end{figure}

\begin{table}[ht]
  \begin{center}
  \caption{Goodness-of-fit of the theoretical and experimental  patterns$^\dag$}\label{tab:Table1}
  \scriptsize
    \begin{tabular}{@{\vrule height 4.5pt depth 4pt  width0pt}lrrrrrrrrrr}
      \hline
\vrule depth 4pt width 0pt &&\multicolumn2c{{\bf $\rho$}}&&&\multicolumn2c{$R^2$}&\\
      \cline{2-9}
 &\multicolumn2c{Unstable
RPS}&\multicolumn2c{Stable RPS}&\multicolumn2c{Unstable
RPS}&\multicolumn2c{Stable
RPS}&\\
Dynamics Model Name &~~~~ $L$ ~~  & ~~~~ $S$ ~~  & ~~~~$L$ ~~  & ~~~~$S$ ~~ &~~~~$L$ ~~   & ~~~~$S$ ~~     & ~~~~ $L$ ~~  & ~~~~ $S$ ~~ \\
\hline
\noalign{\vskip2pt}
Replicator& $ $0.393& $ $0.336& 0.560& 0.479&  0.020& 0.004&  0.086&  $-$0.011\\
BR& 0.844&  0.441&  0.730& 0.483&  0.657&  $-$0.184& 0.251&  $-$2.306\\
MSReplicator&  0.371&  0.315&  0.599&  0.541&  $-$0.005& $-$0.011& 0.151&  0.079\\
BNN&  0.826&  0.853&  0.900&  0.830& 0.679&  0.713&  0.809&  0.688\\
ILogit(0.1)&  0.829&  0.441&  0.726&  0.482&  0.637&  $-$0.121& 0.229&  $-$1.817\\
SampleBR(2) & 0.910& 0.695&  0.820& 0.777&  0.825&  0.464&  0.634&  0.452\\
Smith&  0.736&  0.651&  0.860& 0.848&  0.479&  0.379&  0.694&  0.692\\
Logit(0.001)& 0.844&  0.441&  0.730& 0.483&  0.657&  $-$0.184& 0.251&  $-$2.306\\
Logit(0.01)&  0.844&  0.441&  0.730& 0.483&  0.657&  $-$0.184& 0.251&  $-$2.306\\
Logit(0.1)& 0.844&  0.440& 0.730& 0.483&  0.657&  $-$0.184& 0.251&  $-$2.306\\
Logit(1)& 0.854&  0.512&  0.732&  0.526&  0.676&  $-$0.072& 0.256&  $-$2.030\\
Logit(10)&  0.880& 0.853&  0.841&  0.764&  0.759&  0.720& 0.648&  0.348\\
Logit(100)& 0.681&  0.727&  0.453&  0.563&  0.417&  0.492&  0.201&  $-$0.056\\
Logit(1000)&  0.350& 0.710& 0.153&  0.490& 0.063&  0.459&  0.005&  $-$0.201\\
Projection& $ $0.915& $ $0.859& 0.876&  0.802&  0.837&  0.736&  0.723&  0.557\\
      \hline
    \end{tabular}
  \begin{tablenotes}
  $^\dag$The meaning of the models and their parameters refers to the textbook and its software suite \cite{Sandholm2011,sandholm2007}.
  \end{tablenotes}
  \end{center}
\end{table}

\begin{table}[t]
  \begin{center}
  \caption{Comparisons between dynamics models.}\label{tab:Table2}
  \scriptsize
\begin{tabular}{lrrrrrrrrrrrrrrrrr}
      \hline
  & &$(1)$  &$(2)$  &$(3)$  &$(4)$  &$(5)$  &$(6)$  &$(7)$  &$(8)$  &$(9)$  &$(10)$ &$(11)$ &$(12)$ &$(13)$ &$(14)$ & Score\\
      \hline
(1)&       Replicator&     &      &   &   &   &   &   &   &   &   &   &   &   &   &   $-$5\\
(2)&       BR&  0&    &   &   &   &   &   &   &   &   &   &   &   &   &   $-$6\\
(3)&       MSReplicator&   0&  0&    &   &   &   &   &   &   &   &   &   &   &   &   $-$5\\
(4)&       BNN& 1&  1&  1&    &   &   &   &   &   &   &   &   &   &   &   11\\
(5)&       ILogit(0.1)& 0&  0&  0&   $-1$ &   &   &   &   &   &   &   &   &   &   &   $-$6\\
(6)&       SampleBR(2)& 1&  1&  1&   $-1$ & 1&    &   &   &   &   &   &   &   &   &   8\\
(7)&       Smith&   1&  1&  1&  0&  1&  0&    &   &   &   &   &   &   &   &   9\\
(8)&       Logit(0.001)&    0&  0&  0&   $-1$ & 0&   $-1$ &  $-1$ &   &   &   &   &   &   &   &   $-$6\\
(9)&      Logit(0.01)&  0&  0&  0&   $-1$ & 0&   $-1$ &  $-1$ & 0&    &   &   &   &   &   &   $-$6\\
(10)&     Logit(0.1)&   0&  0&  0&   $-1$ & 0&   $-1$ &  $-1$ & 0&  0&    &   &   &   &   &   $-$6\\
(11)&     Logit(1)& 0&  1&  0&   $-1$ & 1&   $-1$ &  $-1$ & 1&  1&  1&    &   &   &   &   0\\
(12)&     Logit(10)&    1&  1&  1&  0&  1&  0&  0&  1&  1&  1&  1&    &   &   &   9\\
(13)&     Logit(100)&   0&  0&  0&   $-1$ & 0&   $-1$ &  $-1$ & 0&  0&  0&  0&   $-1$ &   &   &   $-$4\\
(14)&     Logit(1000)&  0&  0&  0&   $-1$ & 0&   $-1$ &  $-1$ & 0&  0&  0&  0&   $-1$ &  $-1$ &   &   $-$6\\
(15)&     Projection&   1&  1&  1&  0&  1&  1&  1&  1&  1&  1&  1&  1&  1&  1&   13 \\
      \hline
\end{tabular}
\end{center}
\end{table}


\section{Discussion}

\textbf{Summary ||}
This work makes two contributions to evolutionary game dynamics.
First, by introducing two natural metrics, we illustrate two new geometric patterns, angular momentum and speed (see Fig. \ref{fig:fig2}). To the best of our knowledge, such high-precision empirical dynamic patterns have not been made available in the literature relating to real systems on evolutionary game theory.
Second, based on these geometric patterns, we have illustrated an approach to testing which dynamics models might be more suited to a given game experimental system (see Tables \ref{tab:Table1} and \ref{tab:Table2}).
In previous works (e.g.,\cite{Wang2014Social,Wangxu2014,Friedman2014,Nowak2015,friedman2010tasp}), it had seemed that the dynamics model could be arbitrarily chosen because the models could not be clearly distinguished by experimental observations.
Here, we show that the dynamics model cannot be arbitrarily chosen for a given game experimental system because some typical models are demonstrably unsuitable.
Of course, using our approach, we can also determine which models are appropriate for a given experimental scenario.
Thus, this approach can establish a link between dynamics models and experimental systems, which is, we would argue, by far the most effective and rigorous strategy for the testability of evolutionary game dynamics models.
%
%

\textbf{On comparisons with existing literature ||}
Researchers have clearly shown how to distinguish various games in experiments with models\cite{Wang2014Social,Wangxu2014,Friedman2014,Nowak2015,friedman2010tasp}, but have not successfully shown how to distinguish various models with games experiments. There are probably two reasons for this.
First, the measurements to test the existence of cyclic \cite{XZW2013,Friedman2014,Wangxu2014}, or spiral characteristics for stable (or unstable) games \cite{Friedman2014,Nowak2015}, or the direction of cycling \cite{Friedman2014}, are intended to test the common characteristics of dynamics models, which naturally tends to ignore the differences between the models.
These measurements are therefore not appropriate for capturing the difference between evolutionary dynamics models, although these measurements clearly reveal that dynamics models significantly outperform the Nash equilibrium concept of classical game theory
\cite{XZW2013,Wang2014Social,Wangxu2014,Friedman2014,Nowak2015}.
%
Second, obtaining a deterministic observation from highly stochastic data requires a long time series \cite{Wang2014Social,XZW2013}. When the time series harvested is short, the observed dynamic patterns would be too obscure to reveal distinctions between models
\cite{Friedman2014,Friedman2016,Plott2008,Camerer2003,Berninghaus1999Continuous,Friedman1998,Friedman1996,cason2005dynamics}.
Our measurements, on the other hand, provide sufficient samples from the patches in the geometric patterns (shown in Fig. \ref{fig:fig2}(i--l)), and our approach can evaluate models quantitatively and rigorously (shown in Tables 1 and 2).

\textbf{On further research ||} Although our approach can improve the rigor of merging theory and data, the following questions remain. First, in this study we estimate
models with simple linear regression. We note that the theoretical and experimental values do not match exactly. Indeed, this relates to the open question in evolutionary game theory of methods of normalizing the payoff matrix \cite{Camerer2003,XZW2013,Friedman1991Evolutionary,Friedman2016} or time steps \cite{Friedman2014,Friedman2016}. Although we simply \emph{used the values in the payoff matrix as utility} to specify a model for the game's theoretical patterns, by following existing game literature
\cite{Plott2008,Camerer2003,XZW2013,Wang2014Social,Wangxu2014,Friedman2014,Nowak2015}, we see that it is definitely important to further study normalization \cite{Friedman2016}. Second, in this study, we introduced two observations, $L$ and $S$, though there may be others that can also distinguish the differences among models. For future study, it would be meaningful to consider and include more observations for testing dynamics models (e.g., the distribution in state space \cite{XZW2013,Friedman2014,Nowak2015}), because geometric patterns can provide richer testable samples. Third, we have not compared the empirical patterns (e.g., Fig. 2(i,j) versus Fig. 2(k,l)) for capturing the difference in dynamic behaviors between the stable and unstable systems. This comparison could be a new approach to understanding human social dynamic behaviors, which has been widely studied for decades \cite{Friedman1991Evolutionary,Friedman2014,Nowak2015,Wang2014Social}. Fourth, as macroscopic social dynamics behaviors (called social learning \cite{young2008stochastic}) are rooted in the microscope learning models of games \cite{Borgers1997}, which relate to algorithm game theory and artificial intelligence, our approach can be applied to these fields.
There are other dynamics models involving microscopic update rules (e.g., Moran processes \cite{Sandholm2011} and pairwise imitation \cite{HofbauerSigmund1998}) in addition to the 15 models tested here.
%
Evaluating these models with dynamical observations, like $L$ and $S$ introduced in this paper, could be further tasks.
Fifth, in this study, we have considered only the deterministic motion of time reversal asymmetry. The stochastic motions of time reversal symmetry are ignored, because these do not contribute to the deterministic motions, but these parts can contribute to trajectory distributions and instantaneous motion. It will be necessary to investigate these motions, whose potential methods might relate to recent developments in nonequilibrium statistical physics \cite{XZW2013}. Nevertheless, in summary, we believe that our approach can be generally applied to fields relating to evolutionary game theory and social dynamics (e.g., \cite{Skyrms2014In,Castellano2009,Levin2009Games,Friedman2016,Crockett2013PRICE,Skyrms2014Social}) to rigorously merge theory and experiment.
\appendix
\section{~~~~~~~~~~~~~~Sample resolution}
\label{Ap:SamRe}
In strategy vector \textbf{x}, we know that the three strategy components should satisfy $P_{R} \geq 0$, $P_{P}\geq 0$, $P_{S}\geq 0$, and $P_{R}+P_{P}+P_{S}=1$. These conditions mean that the system states in a triangle can be viewed as located on a triangle plane where $P_{R}+P_{P}+P_{S}=1$. To obtain the sample points for the theoretical models from the triangle plane, we define a resolution parameter $d$ and divide the triangle plane into patches. We assume that in one patch $e_{ij}$, where $i,j \in \mathbb{N}$ and $1 \leq i,j \leq 1/d$, the values of $P_R$ and $P_P$ for one strategy state should satisfy $(i-1)d \leq P_{R} \leq id$ and $(j-1)d \leq P_{P} \leq jd$. For example, when $d=0.25$, the triangle plane is divided into $10$ patches. Let $c_{ij}=[(i-1)d+\frac{d}{2},(j-1)d+\frac{d}{2},1-(i+j-1)d]$, representing the vector for the center point in patch $e_{ij}$. We take the variable value at the center points in each patch of the triangle plane as one theoretical observation, and the observations in all the patches are considered to be theoretical samples for the variable.
In contrast, we compute the average value of all the recorded variable (speed or angular momentum) values in each patch as one experimental observation; the observations in all patches are considered experimental samples for the variable.
In this work, we consider different values for resolution $d$, and found that our results are still valid when the value of the resolution is changed (see Figures S3, S4, S5, S6, S7, S8, S9, S10, and S11 in Supplementary Information).
\section{~~~~~~~~~~~~~~Statistical coefficients and methods}
\label{Ap:rhoR2}
{\bf Pearson correlation coefficient $\rho$.} We used the Pearson correlation coefficient $\rho$ as an index to identify the validity of a model. As a necessary condition, if a model is appropriate for a given experiment, its prediction should be positively linearly correlated with its related experimental observation. The degree of the positive linear correlation can be identified by $\rho$. For a given observation, ideally, $\rho$ should be 1. The larger the value of$\rho$, the more appropriate the model. Alternatively, if $\rho$ is close to 0 or is negative, the model is not appropriate for the experimental system. We used Matlab R2010b to calculate this coefficient.

{\bf Coefficient of determination $R^2$.} $R^2$ indicates the \emph{proportion} of the variance in the dependent variable (experimental observation) that is predictable from the independent variable (theoretical observation).
Assuming that the theoretical observation is expected to be proportional to the experimental observation, our linear regression model contains no constant term. Then, in this study, the value of $R^2$ is not simply the square of the value of $\rho$, and can be negative. For a given observation, ideally, $R^2$ should be 1. If $R^2$ is large, the model is appropriate. Alternatively, if $R^2 $ is close to 0 or is negative, the model is not appropriate for the experimental system. We used Matlab R2010b to calculate this coefficient.


{\bf Statistical methods.} To determine which model performs better of a pair
of evolutionary game dynamics models, we used the Student's t-test ($ttest$) and binomial test ($bitest$). We obtain all possible $\rho$ and $R^2$ values in different cases in which we consider three different resolutions (0.1, 0.05, and 0.025), two RPS games (stable and unstable), and two dynamics observations (speed and the angular momentum).
For the statistical results in each cell in Table 2, the sample points are the signal of the differences of the $\rho$ (or $R^2$) values between the pair. We used Matlab R2010b to report the statistical results.

\section*{Acknowledgments}
This work was partially supported by the Fundamental Research Funds for the Central Universities (SSEYI2014Z) and the National Natural Science Foundation of China (Grant No. 61503062).

\section*{Author Contributions}
%
Y.W. performed theoretical and experimental analyses, Z.W. and X.C. wrote the text, and all authors designed the research and approved this manuscript.

\section*{Competing Financial Interests}
The authors declare no competing financial interests.



\begin{thebibliography}{00}


\bibitem{VonNeumann1944}
\bibinfo{author}{Von~Neumann, J.} \& \bibinfo{author}{Morgenstern, O.}
\newblock \emph{\bibinfo{title}{Theory of games and economic behavior}}
  (\bibinfo{publisher}{Princeton University Press}, \bibinfo{year}{1944}).

\bibitem{myerson1997game}
\bibinfo{author}{Myerson, R.}
\newblock \emph{\bibinfo{title}{Game theory: analysis of conflict}}
  (\bibinfo{publisher}{Harvard Univ Pr}, \bibinfo{year}{1997}).

\bibitem{darwin1859origin}
\bibinfo{author}{Darwin, C.}
\newblock \emph{\bibinfo{title}{On the origin of species}}
  (\bibinfo{publisher}{Murray, London}, \bibinfo{year}{1859}).

\bibitem{Smith1982}
\bibinfo{author}{Maynard~Smith, J.}
\newblock \emph{\bibinfo{title}{Evolution and the theory of games}}
  (\bibinfo{publisher}{Cambridge University Press}, \bibinfo{year}{1982}).

\bibitem{Levin2009Games}
\bibinfo{author}{Levin, S.~A.}
\newblock \emph{\bibinfo{title}{Games, groups, and the global good}}
  (\bibinfo{publisher}{Springer}, \bibinfo{year}{2009}).

\bibitem{Weibull1997}
\bibinfo{author}{Weibull, J.}
\newblock \emph{\bibinfo{title}{Evolutionary game theory}}
  (\bibinfo{publisher}{The MIT Press}, \bibinfo{year}{1997}).

\bibitem{HofbauerSigmund1998}
\bibinfo{author}{Hofbauer, J.} \& \bibinfo{author}{Sigmund, K.}
\newblock \emph{\bibinfo{title}{Evolutionary games and population dynamics}}
  (\bibinfo{publisher}{Cambridge University Press}, \bibinfo{year}{1998}).

\bibitem{Skyrms2014Social}
\bibinfo{author}{Skyrms, B.}
\newblock \emph{\bibinfo{title}{Social dynamics}} (\bibinfo{publisher}{Oxford University Press}, \bibinfo{year}{2014}).

\bibitem{Samuelson2002}
\bibinfo{author}{Samuelson, L.}
\newblock \bibinfo{title}{Evolution and game theory}.
\newblock \emph{\bibinfo{journal}{The Journal of Economic Perspectives}}
  \textbf{\bibinfo{volume}{16}}, \bibinfo{pages}{47--66}
  (\bibinfo{year}{2002}).

\bibitem{Skyrms2014In}
\bibinfo{author}{Skyrms, B.}, \bibinfo{author}{Avise, J.~C.} \&
  \bibinfo{author}{Ayala, F.~J.}
\newblock \bibinfo{title}{In the light of evolution viii: Darwinian thinking in
  the social sciences.}
\newblock \emph{\bibinfo{journal}{Proceedings of the National Academy of
  Sciences of the United States of America}} \textbf{\bibinfo{volume}{111 Suppl 3}}, \bibinfo{pages}{10781--10784} (\bibinfo{year}{2014}).

\bibitem{Friedman1998Rev}
\bibinfo{author}{Friedman, D.}
\newblock \bibinfo{title}{Evolutionary economics goes mainstream: A review of the theory of learning in games}.
\newblock \emph{\bibinfo{journal}{Journal of Evolutionary Economics}}
  \textbf{\bibinfo{volume}{8}}, \bibinfo{pages}{423--432}
  (\bibinfo{year}{1998}).

\bibitem{nowak2006evolutionary}
\bibinfo{author}{Nowak, M.~A.}
\newblock \emph{\bibinfo{title}{Evolutionary dynamics: Exploring the equations of life}} (\bibinfo{publisher}{Harvard University Press},
  \bibinfo{year}{2006}).

\bibitem{Sandholm2011}
\bibinfo{author}{Sandholm, W.}
\newblock \emph{\bibinfo{title}{Population games and evolutionary dynamics}}
  (\bibinfo{publisher}{The MIT Press}, \bibinfo{year}{2011}).

\bibitem{frey2010evolutionary}
\bibinfo{author}{Frey, E.}
\newblock \bibinfo{title}{Evolutionary game theory: Theoretical concepts and applications to microbial communities}.
\newblock \emph{\bibinfo{journal}{Physica A: Statistical Mechanics and its Applications}} \textbf{\bibinfo{volume}{389}}, \bibinfo{pages}{4265--4298}
  (\bibinfo{year}{2010}).

\bibitem{Taylor1978}
\bibinfo{author}{Taylor, P.} \& \bibinfo{author}{Jonker, L.}
\newblock \bibinfo{title}{Evolutionary stable strategies and game dynamics}.
\newblock \emph{\bibinfo{journal}{Mathematical Biosciences}}
  \textbf{\bibinfo{volume}{40}}, \bibinfo{pages}{145--156}
  (\bibinfo{year}{1978}).

\bibitem{sandholm201603}
\bibinfo{author}{Mertikopoulos, P.} \& \bibinfo{author}{Sandholm, W.}
\newblock \bibinfo{title}{Riemannian game dynamics}.
\newblock \emph{\bibinfo{journal}{Arxiv preprint}}
  \textbf{\bibinfo{volume}{1603.09173}} (\bibinfo{year}{2016}).


\bibitem{Friedman2014}
\bibinfo{author}{Cason, T.~N.}, \bibinfo{author}{Friedman, D.} \&
  \bibinfo{author}{Hopkins, E.}
\newblock \bibinfo{title}{Cycles and instability in a rock-paper-scissors population game: a continuous time experiment}.
\newblock \emph{\bibinfo{journal}{Review of Economic Studies}}
  \textbf{\bibinfo{volume}{81}}, \bibinfo{pages}{112--136}
  (\bibinfo{year}{2014}).

\bibitem{Crawford1991}
\bibinfo{author}{Crawford, V.}
\newblock \bibinfo{title}{An evolutionary interpretation of van Huyck, Battalio, and Beil's experimental results on coordination}.
\newblock \emph{\bibinfo{journal}{Games and Economic Behavior}}
  \textbf{\bibinfo{volume}{3}}, \bibinfo{pages}{25--59} (\bibinfo{year}{1991}).

\bibitem{Miller1991Can}
\bibinfo{author}{Miller, J.~H.} \& \bibinfo{author}{Andreoni, J.}
\newblock \bibinfo{title}{Can evolutionary dynamics explain free riding in experiments?}
\newblock \emph{\bibinfo{journal}{Economics Letters}}
  \textbf{\bibinfo{volume}{36}}, \bibinfo{pages}{9--15} (\bibinfo{year}{1991}).

\bibitem{Plott2001Equilibrium}
\bibinfo{author}{Plott, C.~R.}
\newblock \bibinfo{title}{Equilibrium, equilibration, information and multiple markets: From basic science to institutional design}.
\newblock \emph{\bibinfo{journal}{Nobel Symposium Behavioral \& Experimental Economics}}  (\bibinfo{year}{2001}).

\bibitem{Plott2008}
\bibinfo{author}{Plott, C.} \& \bibinfo{author}{Smith, V.}
\newblock \emph{\bibinfo{title}{Handbook of experimental economics results}}
  (\bibinfo{publisher}{North-Holland}, \bibinfo{year}{2008}).

\bibitem{Friedman1996}
\bibinfo{author}{Friedman, D.}
\newblock \bibinfo{title}{Equilibrium in evolutionary games: Some experimental results}.
\newblock \emph{\bibinfo{journal}{The Economic Journal}}
  \textbf{\bibinfo{volume}{106}}, \bibinfo{pages}{1--25}
  (\bibinfo{year}{1996}).

\bibitem{Wang2014Social}
\bibinfo{author}{Wang, Z.}, \bibinfo{author}{Xu, B.} \& \bibinfo{author}{Zhou, H.~J.}
\newblock \bibinfo{title}{Social cycling and conditional responses in the
  rock-paper-scissors game.}
\newblock \emph{\bibinfo{journal}{Scientific Reports}}
  \textbf{\bibinfo{volume}{4}}, \bibinfo{pages}{5830--5830}
  (\bibinfo{year}{2014}).

\bibitem{Nowak2015}
\bibinfo{author}{Hoffman, M.}, \bibinfo{author}{Suetens, S.},
  \bibinfo{author}{Nowak, M.} \& \bibinfo{author}{Gneezy, U.}
\newblock \bibinfo{title}{An experimental investigation of evolutionary
  dynamics in the rock-paper-scissors game}.
\newblock \emph{\bibinfo{journal}{Scientific Reports}}
  \textbf{\bibinfo{volume}{5}}, \bibinfo{pages}{8817} (\bibinfo{year}{2015}).

\bibitem{Friedman2016}
\bibinfo{author}{Friedman, D.} \& \bibinfo{author}{Sinervo, B.}
\newblock \emph{\bibinfo{title}{Evolutionary games in natural, social, and
  virtual worlds}} (\bibinfo{publisher}{Oxford University Press},
  \bibinfo{year}{2016}).

\bibitem{Wangxu2014}
\bibinfo{author}{Wang, Z.} \& \bibinfo{author}{Xu, B.}
\newblock \bibinfo{title}{Incentive and stability in the rock-paper-scissors game: An experimental investigation}.
\newblock \emph{\bibinfo{journal}{Arxiv preprint arXiv:1407.1170}}
  (\bibinfo{year}{2010}).

\bibitem{Sin96}
\bibinfo{author}{Sinervo, B.} \& \bibinfo{author}{Lively, C.}
\newblock \bibinfo{title}{The rock-paper-scissors game and the evolution of
  alternative male strategies}.
\newblock \emph{\bibinfo{journal}{Nature}} \textbf{\bibinfo{volume}{380}},
  \bibinfo{pages}{240--243} (\bibinfo{year}{1996}).

\bibitem{pryke2006red}
\bibinfo{author}{Pryke, S.~R.} \& \bibinfo{author}{Griffith, S.~C.}
\newblock \bibinfo{title}{Red dominates black: agonistic signalling among head morphs in the colour polymorphic Gouldian finch}.
\newblock \emph{\bibinfo{journal}{Proceedings of the Royal Society B:
  Biological Sciences}} \textbf{\bibinfo{volume}{273}},
  \bibinfo{pages}{949--957} (\bibinfo{year}{2006}).

\bibitem{Ker02}
\bibinfo{author}{Kerr, B.}, \bibinfo{author}{Riley, M.},
  \bibinfo{author}{Feldman, M.} \& \bibinfo{author}{Bohannan, B.}
\newblock \bibinfo{title}{Local dispersal promotes biodiversity in a real-life game of rock-paper-scissors}.
\newblock \emph{\bibinfo{journal}{Nature}} \textbf{\bibinfo{volume}{418}},
  \bibinfo{pages}{171--174} (\bibinfo{year}{2002}).

\bibitem{Levy2015Quantitative}
\bibinfo{author}{Levy, S.~F.} \emph{et~al.}
\newblock \bibinfo{title}{Quantitative evolutionary dynamics using high-resolution lineage tracking.}
\newblock \emph{\bibinfo{journal}{Nature}} \textbf{\bibinfo{volume}{519}},
  \bibinfo{pages}{181--186} (\bibinfo{year}{2015}).

\bibitem{friedman2010tasp}
\bibinfo{author}{Cason, T.~N.}, \bibinfo{author}{Friedman, D.} \&
  \bibinfo{author}{Hopkins, E.}
\newblock \bibinfo{title}{Testing the tasp: An experimental investigation of learning in games with unstable equilibria}.
\newblock \emph{\bibinfo{journal}{Journal of Economic Theory}}
  \textbf{\bibinfo{volume}{145}}, \bibinfo{pages}{2309--2331}
  (\bibinfo{year}{2010}).

\bibitem{Zhou2016}
\bibinfo{author}{Zhou, H.-J.}
\newblock \bibinfo{title}{The rock-paper-scissors game}.
\newblock \emph{\bibinfo{journal}{Contemporary Physics}}
  \textbf{\bibinfo{volume}{57}}, \bibinfo{pages}{151--163}
  (\bibinfo{year}{2016}).

\bibitem{XZW2013}
\bibinfo{author}{{Xu}, B.}, \bibinfo{author}{{Zhou}, H.-J.} \&
  \bibinfo{author}{{Wang}, Z.}
\newblock \bibinfo{title}{{Cycle frequency in standard rock-paper-scissors games: Evidence from experimental economics}}.
\newblock \emph{\bibinfo{journal}{Physica A: Statistical Mechanics and its Applications}} \textbf{\bibinfo{volume}{392}}, \bibinfo{pages}{4997--5005}
  (\bibinfo{year}{2013}).

\bibitem{Xu2014}
\bibinfo{author}{Xu, B.}, \bibinfo{author}{Wang, S.} \& \bibinfo{author}{Wang, Z.}
\newblock \bibinfo{title}{Periodic frequencies of the cycles in 2 $\times$ 2 games: evidence from experimental economics}.
\newblock \emph{\bibinfo{journal}{The European Physical Journal B}}
  \textbf{\bibinfo{volume}{87}}, \bibinfo{pages}{1--10} (\bibinfo{year}{2014}).

\bibitem{Wang201009}
\bibinfo{author}{Wang, Z.}
\newblock \bibinfo{title}{Social spiral pattern in experimental 2x2 games}.
\newblock \emph{\bibinfo{journal}{Arxiv preprint arXiv:1009.3560}}
  (\bibinfo{year}{2010}).

\bibitem{XuWang2011ICCS}
\bibinfo{author}{Xu, B.} \& \bibinfo{author}{Wang, Z.}
\newblock \emph{\bibinfo{title}{Evolutionary dynamical pattern of "coyness and philandering": Evidence from experimental economics}}, vol.
  \bibinfo{volume}{VIII} (\bibinfo{publisher}{p1313-1326, NECSI Knowledge
  Press, ISBN 978-0-9656328-4-3.}, \bibinfo{year}{2011}).

\bibitem{xu2012test}
\bibinfo{author}{Xu, B.} \& \bibinfo{author}{Wang, Z.}
\newblock \bibinfo{title}{Test maxent in social strategy transitions with
  experimental two-person constant sum $2 \times 2$ games}.
\newblock \emph{\bibinfo{journal}{Results in Physics}}
  \textbf{\bibinfo{volume}{2}},
  \bibinfo{pages}{127--134}
  (\bibinfo{year}{2012}).

\bibitem{Friedman2011Separating}
\bibinfo{author}{Oprea, R.}, \bibinfo{author}{Henwood, K.} \&
  \bibinfo{author}{Friedman, D.}
\newblock \bibinfo{title}{Separating the hawks from the doves: Evidence from continuous time laboratory games}.
\newblock \emph{\bibinfo{journal}{Journal of Economic Theory}}
  \textbf{\bibinfo{volume}{146}}, \bibinfo{pages}{2206--2225}
  (\bibinfo{year}{2011}).

\bibitem{sandholm2007}
\bibinfo{author}{Sandholm, W.} \& \bibinfo{author}{Dokumaci, E.}
\newblock \bibinfo{title}{Dynamo: Phase diagrams for evolutionary dynamics}.
\newblock \emph{\bibinfo{journal}{Software suite}}  (\bibinfo{year}{2007}).

\bibitem{selten2008}
\bibinfo{author}{Selten, R.} \& \bibinfo{author}{Chmura, T.}
\newblock \bibinfo{title}{Stationary concepts for experimental $2 \times 2$-games}.
\newblock \emph{\bibinfo{journal}{The American Economic Review}}
  \textbf{\bibinfo{volume}{98}}, \bibinfo{pages}{938--966}
  (\bibinfo{year}{2008}).

\bibitem{RothErev2007}
\bibinfo{author}{Erev, I.}, \bibinfo{author}{Roth, A.},
  \bibinfo{author}{Slonim, R.} \& \bibinfo{author}{Barron, G.}
\newblock \bibinfo{title}{Learning and equilibrium as useful approximations: Accuracy of prediction on randomly selected constant sum games}.
\newblock \emph{\bibinfo{journal}{Economic Theory}}
  \textbf{\bibinfo{volume}{33}}, \bibinfo{pages}{29--51}
  (\bibinfo{year}{2007}).

\bibitem{Camerer2003}
\bibinfo{author}{Camerer, C.} \& \bibinfo{author}{Foundation, R.~S.}
\newblock \emph{\bibinfo{title}{Behavioral game theory: Experiments in
  strategic interaction}}, vol.~\bibinfo{volume}{9}
  (\bibinfo{publisher}{Princeton University Press Princeton, NJ},
  \bibinfo{year}{2003}).

\bibitem{Berninghaus1999Continuous}
\bibinfo{author}{Berninghaus, S.~K.}, \bibinfo{author}{K.-M., E.} \&
  \bibinfo{author}{Keser, C.}
\newblock \bibinfo{title}{Continuous-time strategy selection in linear
  population games}.
\newblock \emph{\bibinfo{journal}{Experimental Economics}}
  \textbf{\bibinfo{volume}{2}}, \bibinfo{pages}{41--57} (\bibinfo{year}{1999}).

\bibitem{Friedman1998}
\bibinfo{author}{Cheung, Y.} \& \bibinfo{author}{Friedman, D.}
\newblock \bibinfo{title}{A comparison of learning and replicator dynamics
  using experimental data}.
\newblock \emph{\bibinfo{journal}{Journal of Economic Behavior and
  Organization}} \textbf{\bibinfo{volume}{35}}, \bibinfo{pages}{263--280}
  (\bibinfo{year}{1998}).

\bibitem{cason2005dynamics}
\bibinfo{author}{Cason, T.}, \bibinfo{author}{Friedman, D.} \&
  \bibinfo{author}{Wagener, F.}
\newblock \bibinfo{title}{{The dynamics of price dispersion, or Edgeworth
  variations}}.
\newblock \emph{\bibinfo{journal}{Journal of Economic Dynamics and Control}}
  \textbf{\bibinfo{volume}{29}}, \bibinfo{pages}{801--822}
  (\bibinfo{year}{2005}).

\bibitem{Friedman1991Evolutionary}
\bibinfo{author}{Friedman, D.}
\newblock \bibinfo{title}{Evolutionary games in economics}.
\newblock \emph{\bibinfo{journal}{Econometrica}} \textbf{\bibinfo{volume}{59}},
  \bibinfo{pages}{637--66} (\bibinfo{year}{1991}).

\bibitem{young2008stochastic}
\bibinfo{author}{Young, H.}
\newblock \bibinfo{title}{{Stochastic adaptive dynamics}}.
\newblock \emph{\bibinfo{journal}{New Palgrave Dictionary of Economics, revised edition, L. Blume and S. Durlauf, eds. Zanella, G.(2007), Discrete Choice with Social Interactions and Endogenous Membership, Journal of the European Economic Association}} \textbf{\bibinfo{volume}{5}},
  \bibinfo{pages}{122--153} (\bibinfo{year}{2008}).

\bibitem{Borgers1997}
\bibinfo{author}{Borgers, T.} \& \bibinfo{author}{Sarin, R.}
\newblock \bibinfo{title}{Learning through reinforcement and replicator
  dynamics}.
\newblock \emph{\bibinfo{journal}{Journal of Economic Theory}}
  \textbf{\bibinfo{volume}{77}}, \bibinfo{pages}{1--14} (\bibinfo{year}{1997}).

\bibitem{Castellano2009}
\bibinfo{author}{Castellano, C.} \& \bibinfo{author}{Loreto, V.}
\newblock \bibinfo{title}{Statistical physics of social dynamics}.
\newblock \emph{\bibinfo{journal}{Reviews of Modern Physics}}
  \textbf{\bibinfo{volume}{81}}, \bibinfo{pages}{591--646} (\bibinfo{year}{2009}).

\bibitem{Crockett2013PRICE}
\bibinfo{author}{Crockett, S.}
\newblock \bibinfo{title}{Price dynamics in general equilibrium experiments}.
\newblock \emph{\bibinfo{journal}{Journal of Economic Surveys}}
  \textbf{\bibinfo{volume}{27}}, \bibinfo{pages}{421--438}
  (\bibinfo{year}{2013}).

\end{thebibliography}
\end{document}